\documentclass[%
 reprint, superscriptaddress,
 amsmath,amssymb,
 aps,
]{revtex4-2}
\usepackage{graphicx}% Include figure files
\usepackage{xcolor}
\usepackage{physics}
\usepackage{amsfonts,mathtools}
\usepackage{amsthm}
\usepackage{CJKutf8}
\usepackage{physics,dsfont}
\usepackage{multirow}
\usepackage{graphicx}% Include figure files
\usepackage{dcolumn}% Align table columns on decimal point
\usepackage{bm}% bold math
\usepackage{hyperref}% add hypertext capabilities
\newcommand{\mL}{\mathcal{L}}
\newcommand{\mQ}{\mathcal{Q}}
\newcommand{\mC}{\mathcal{L}_{+1}}
\newcommand{\mA}{\mathcal{A}}
\newcommand{\mB}{\mathcal{B}}
\newcommand{\mX}{\mathcal{X}}
\newcommand{\mY}{\mathcal{Y}}
\newcommand{\mNS}{\mathcal{NS}}

\newcommand{\revise}[1]{{\color{black}#1}}

\begin{document}

\title{
\begin{CJK*}{UTF8}{gbsn}
Beating one bit of communication with quantum correlations in smaller dimensions% Force line breaks with \\
\end{CJK*}
}

\author{Peter Sidajaya}
\email{peter.sidajaya@u.nus.edu}
\affiliation{Centre for Quantum Technologies, National University of Singapore, 3 Science Drive 2, Singapore 117543}

\author{Valerio Scarani}
\affiliation{Centre for Quantum Technologies, National University of Singapore, 3 Science Drive 2, Singapore 117543}
\affiliation{Department of Physics, National University of Singapore, 2 Science Drive 3, Singapore 117542}

\date{\today}% It is always \today, today,
             %  but any date may be explicitly specified

\begin{abstract}
As a consequence of Bell's theorem, the statistics of measurements on some entangled states cannot be simulated with local hidden variables alone. The amount of communication that must be supplied is an intuitive quantifier of nonclassicality. While it is obvious that this amount can be very large in general, it has been surprisingly difficult to find simple examples of quantum correlations, whose simulation requires more than one bit of communication. In this paper, we report the simplest example to date, which lives in the $(5,2,5,5)$ Bell scenario [the previously known smallest case living in the $(7,3,16,16)$ scenario]. The proof is built on the observation that finding the largest 1-bit score is equivalent to finding the bipartition of the inputs, in which the sum of the local scores of the two subgames is maximal.
\end{abstract}

%\keywords{Suggested keywords}%Use showkeys class option if keyword
                              %display desired
\maketitle

\section{Introduction}

Since Bell's seminal theorem, we have known that the correlations made by some quantum systems cannot be reproduced solely with local hidden variables (LHVs) \cite{bell1964einstein,scarani2019bell}. Consequently, some have started asking how powerful are these quantum correlations in contrast to LHV, and compared to one another. One way to phrase this question is to ask how many additional resources are needed to supplement LHV in order to simulate these quantum correlations. Some have investigated the case for \textit{nonlocal boxes} \cite{popescu1994quantum, brunner2005entanglement,brunner2008simulation}, but one resource that is most intuitive is classical communication, which we will be focusing on in this paper.

Some work has been done on the average amount of communication needed to simulate the entanglement of a system of a generic dimension \cite{brassard1999cost,massar2001classical,degorre2007classical,brassard2019remote}. On the other hand, going to a small, finite dimension, a substantial body of literature has been written on what could and could not be simulated with just a single bit of classical communication (1 bit). For two qubits, 1 bit is provably enough to simulate all the statistics of the maximally entangled state \cite{brassard1999cost, steiner2000towards, csirik2002cost,toner2003communication,degorre2005simulating} and of a whole range of weakly entangled states \cite{renner2022minimal}; there is numerical evidence supporting the conjecture that all two-qubit states can be simulated with just 1 bit \cite{sidajaya2023neural}. \revise{Of course, the correlations achievable with LHVs supplemented by one bit go beyond two-qubit quantum correlations \cite{PhysRevA.72.050302}.}

Conversely, some have tried to search for a correlation that could not be simulated by 1 bit. These correlations, or \textit{behaviors}, can be sought for in Bell scenarios with $m_{A/B}$ inputs and $o_{A/B}$ outputs for Alice/Bob, denoted $(m_A,m_B,o_A,o_B)$. \revise{In a typical problem in Bell nonlocality, we are usually only interested in the local set, which forms a polytope in the full probability space, and the quantum set, which due to the no-signaling constraint both lie in a subspace of the full probability space called the no-signaling space. In our problem, the set of 1 bit correlations also forms a polytope, akin to the local set, but obviously includes signaling behaviors and thus spans outside the no-signaling space. The real dimension of this full probability space, which we will call the \textit{probability dimension}, is given by $m_Am_Bo_Ao_B$} \footnote{Taking normalization into account, the effective probability dimension is $m_Am_B(o_Ao_B-1)$. We keep the formula in the main text to facilitate direct comparison with Ref.~\cite{marton2023beating}.}. Some have characterized these polytopes in the simplest scenarios, i.e.~up to $(3,3,2,2)$, and found no violation by quantum correlations \cite{bacon2003bell,maxwell2014bell,zambrini2019bell}. For larger scenarios, the task becomes computationally unfeasible. Using a different technique, Vértesi and Bene found a quantum correlation that is not simulatable with 1 bit, although it requires an infinite number of settings \cite{PhysRevA.80.062316}. Eventually, just recently, Márton \textit{et al.}~found examples of quantum correlations that are not simulatable with 1 bit in finite scenarios \cite{marton2023beating}. Their constructions are based on parallel repetition. The smallest scenario in which they found examples are $(7,3,16,16)$ (probability dimension 5376) for fixed-directional communication, and $(7,7,16,16)$ (probability dimension 12544) or $(63,63,2,2)$ (probability dimension 15876) for bidirectional communication. In this paper, we will present an example of a quantum correlation that is not simulatable with 1 bit for fixed directional communication, in the $(5,2,5,5)$ scenario (probability dimension 250). The statistics of interest can be generated by suitable measurements on the maximally entangled state of two five-dimensional systems. This comes within the range of challenging but feasible experimental demonstrations \cite{erhard2020advances,mair2001entanglement,leach2010quantum,dada2011experimental,krenn2014generation,erhard2018twisted, rossi2009multipath, schaeff2012scalable, schaeff2015experimental, wang2018multidimensional, stucki2005fabry, ikuta2016enhanced,bernhard2013shaping, jin2016simple, thew2004bell, richart2012experimental, hrmo2023native}.

The paper is structured as follows. In Sec.~\ref{sec:bell} we will first present a method to find the 1 bit bound that is faster than a brute force approach. In Sec.~\ref{sec:example} we will present our Bell inequality and the corresponding correlation that lies outside the 1 bit polytope.

\section{Bell inequalities for one bit of communication}
\label{sec:bell}

\subsection{Defining the 1-bit polytope}

% \valerio{I had to introduce the notation for the Bell scenario above. I let you decide whether you want to change before, or change below.}

Let us call the local set $\mL$, the quantum set $\mQ$, and the set of 1 bit communication strategies $\mC$. Similar to $\mL$, $\mC$ is also a polytope, but one that spans outside the no-signaling space. Since it is a polytope, the maximum value of a linear inequality achieved by the set must be given by at least one of its extremum points. This simplifies the problem of finding the 1 bit bound into finding the extremum point which maximizes the inequality.

Now let us call the set of inputs of Alice (Bob) as $\mX \; (\mY)$ with size $m_A$ ($m_B$) and her (his) outputs $\mA \; (\mB)$ with size $o_A$ ($o_B$). The specific scenario of a game or inequality is denoted by the $(m_A,m_B,o_A,o_B)$ notation. In the usual studies of Bell nonlocality, the nonsignaling constraints on $\mL$ effectively reduce the dimension of the probability space under study to $m_A(o_A-1)m_B(o_B-1) + m_A(o_A-1) + m_B(o_B-1)$ \cite{collins2004relevant}. The polytope $\mC$, however, is not restricted by this constraint: We must work in the full probability space of dimension $m_Am_Bo_Ao_B$.

After receiving her input, Alice has $o_A^{m_A}$ possible deterministic strategies. She then has $\left(2^{m_A-1}-1\right)$ deterministic communication strategies. This number is given by Stirling's number of the second kind, which gives the number of ways to partition a set into a fixed number of subsets. Finally, Bob has $o_B^{2m_B}$ possible deterministic strategies, where the two come from Alice's bit of communication. This number can be reduced as there are duplicates, with the final number being
$$
o_A^{m_A}\left[
o_B^{m_B} + (2^{m_A-1}-1)(o_B^{2m_B}-o_B^{m_B})
\right].
$$

The number of extremum points grows very quickly with dimensions and direct facet enumeration is practically impossible once we go to $(3,3,2,2)$, where some ingenious techniques are needed to enumerate the facets \cite{zambrini2019bell}. \revise{However, if we are given a Bell inequality, we could still find its 1 bit bound by calculating the value of the Bell inequality on all the extremum points using brute force. Once we go above 20 million points or so, however, which is reached in $(4,3,4,4)$, this also starts becoming unwieldy. This problem can be circumvented by noting that in order to find the maximum 1 bit bound of a Bell inequality, we do not need to find the value of the Bell inequality of every single extremum point of $\mC$.}

\subsection{A faster computation for the 1 bit bound}

A behavior in a given scenario is given by a set of probabilities $p(a,b|x,y)$ for all $a,b,x,y$, which can be written as a vector. Similarly, a Bell inequality is given by a set of coefficients $V(a,b|x,y)$ for all $a,b,x,y$, which can also be written as a vector. As mentioned, the maximum score achieved by a polytope is obtained by at least one of its extremum points. Thus, the task of finding the 1 bit bound of a given Bell inequality $\Vec{B}$ can be written as
\begin{align}
\label{eq:optimize}
    \max_{\Vec{p}\in\{\Vec{p}_{\mC,D}\}}{\Vec{B}\cdot\Vec{p}},
\end{align}
where $\{\Vec{p}_{\mC,D}\}$ is the set of extremum points of $\mC$. Each of the extremum points corresponds to a deterministic strategy. For each point, the output functions of Alice and Bob are given by $A(x)=\delta_{f(x),a}$ and $B(y,c)=\delta_{g(y,c),b}$, while the communication function is given by $C(x)=\delta_{h(x),c}$, for some assignment functions $f,g,h$ that assigns one value $a,b,c$ to every input combinations.

In a deterministic 1 bit communication strategy, Alice deterministically sends a 0 or 1 to Bob depending on her input. This means that Alice and Bob have agreed on partitioning Alice's inputs in two sets,
\begin{align}\label{eq:partition}
    \mathcal{J} = \{x\in\mX | C(x)=0 \}
\end{align}
and its complement $\mX/\mathcal{J}$. When we consider the partitioned sub-strategies, they are local, as we have used the bit of communication. Thus, all deterministic 1 bit communication strategies can be decomposed into two local parts,
\begin{align}\label{eq:p direct sum}
    \Vec{p}_{\mC,D} = \Vec{p}_{\mL',D}^{\;\mathcal{J}} \oplus \Vec{p}_{\mL'',D}^{\;\mX/\mathcal{J}},
\end{align}
where $\oplus$ is the direct addition, $\mL'$ is the local polytope in the smaller scenario of $(|\mathcal{J}|,m_B,o_A,o_B)$ and similarly for $\mL''$. The two substrategies are local in their respective partitions and thus are extremum points of the local polytope in the smaller scenarios. Similarly, we can also partition $\Vec{B}$ into two:
\begin{align}\label{eq:B direct sum}
    \Vec{B} = \Vec{B}^{\;\mathcal{J}} \oplus \Vec{B}^{\;\mX/\mathcal{J}}.
\end{align}

With these in mind, we can rewrite Eq.~(\ref{eq:optimize}) in the following way:
\begin{widetext}
\begin{align}
    \max_{\Vec{p}\in\{\Vec{p}_{\mC,D}\}}{\Vec{B}\cdot\Vec{p}} &= \max_{A(x),B(y,c),C(x)}{\Vec{B}\cdot\Vec{p}_{\mC,D}} \nonumber \\
    &= \max_{A(x),B(y,c),C(x)}{(\Vec{B}^{\;\mathcal{J}} \oplus \Vec{B}^{\;\mX/\mathcal{J}})\cdot(\Vec{p}_{\mL',D}^{\;\mathcal{J}} \oplus \Vec{p}_{\mL'',D}^{\;\mX/\mathcal{J}})} \nonumber \\
    &= \max_{A(x),B(y,c),C(x)}\left[({\Vec{B}^{\;\mathcal{J}}\cdot\Vec{p}_{\mL',D}^{\;\mathcal{J}}) + (\Vec{B}^{\;\mX/\mathcal{J}}\cdot\Vec{p}_{\mL'',D}^{\;\mX/\mathcal{J}})}\right] \nonumber \\
    &= \max_{C(x)}\left[\max_{A'(x),B(y,0)}({\Vec{B}^{\;\mathcal{J}}\cdot\Vec{p}_{\mL',D}^{\;\mathcal{J}}) + \max_{A''(x),B(y,1)}(\Vec{B}^{\;\mX/\mathcal{J}}\cdot\Vec{p}_{\mL'',D}^{\;\mX/\mathcal{J}})}\right].
\end{align}
\end{widetext}
In the first equality, we rewrite the maximization variables to be the strategies of the parties. In the second equality, we use Eqs.~(\ref{eq:p direct sum}) and (\ref{eq:B direct sum}). In the third equality, we partition the Bell vector along $\mathcal{J}$ and use the distributive property of the dot product and the direct sum. In the final equality, we note that the two terms are independent of each other, i.e., the local strategy of the first subinequality does not affect the score of the second sub-inequality. Thus, for a given partition $\mathcal{J}$, we can split the maximization into two terms. Here, we write $A'$ and $A''$ the strategies of Alice to denote that they are defined on the domain of the partitioned input sets. We can now also assign a value to the $c$ in the input arguments of $B$ since $c$ has a fixed value in each of the partitions as defined by Eq.~(\ref{eq:partition}).

For a more intuitive description, consider that for a given Bell game, a deterministic bit of communication allows the parties to effectively partition the game into two: one where her input falls in $\mathcal{J}$, and one where it does not. When she sends the bit to Bob, Bob now knows which partition of the game her input is in. In each of these partitions, they now play a normal Bell game. These two local strategies of the subgames do not interact with each other, since the strategy in the first subgame does not affect how they play and how they score in the second subgame. Thus, a 1 bit strategy is optimum in the total Bell game \textit{if and only if} its sub-strategies are also optimum in the subgames. Thus, in order to find the maximum 1 bit strategy, one needs only to maximize the two subgames using local deterministic strategies, for all the possible ways to partition the Bell game into two subgames.

This rewritten maximization can be run faster than the original form since finding the local bound of a game is a relatively fast task compared to enumerating over all 1 bit deterministic strategies. This speedup comes from the fact that when we rewrite this optimization, we skip over the extremum points which are not optimum in the partitioned subgames defined by the communication strategies. However, note that the result of the optimization is still analytically \textit{exact}, and this allows us to find the \textit{exact} 1 bit bound in higher dimensions, which will be useful for the next sections.

\section{Finding quantum correlations that violate $\mC$}
\label{sec:example}
\subsection{Five (and beyond)-dimensional Bell-like inequalities for $\mC$}
\label{subsec:five-dimension}

\begin{table}[h!]
\centering
\begin{tabular}{c|ccccc|ccccc|}
\cline{2-11}
                                         & \multicolumn{5}{c|}{$y=1$} & \multicolumn{5}{c|}{$y=2$} \\ \hline
\multicolumn{1}{|l|}{\multirow{5}{*}{$x=1$}} & 1 & $\cdot$ & $\cdot$ & $\cdot$ & $\cdot$ & 1 & $\cdot$ & $\cdot$ & $\cdot$ & $\cdot$ \\
\multicolumn{1}{|l|}{} & $\cdot$ & 1 & $\cdot$ & $\cdot$ & $\cdot$ & $\cdot$ & 1 & $\cdot$ & $\cdot$ & $\cdot$ \\
\multicolumn{1}{|l|}{} & $\cdot$ & $\cdot$ & 1 & $\cdot$ & $\cdot$ & $\cdot$ & $\cdot$ & 1 & $\cdot$ & $\cdot$ \\
\multicolumn{1}{|l|}{} & $\cdot$ & $\cdot$ & $\cdot$ & 1 & $\cdot$ & $\cdot$ & $\cdot$ & $\cdot$ & 1 & $\cdot$ \\
\multicolumn{1}{|l|}{} & $\cdot$ & $\cdot$ & $\cdot$ & $\cdot$ & 1 & $\cdot$ & $\cdot$ & $\cdot$ & $\cdot$ & 1 \\ \hline

\multicolumn{1}{|l|}{\multirow{5}{*}{$x=2$}} & 1 & $\cdot$ & $\cdot$ & $\cdot$ & $\cdot$ & $\cdot$ & 1 & $\cdot$ & $\cdot$ & $\cdot$ \\
\multicolumn{1}{|l|}{} & $\cdot$ & 1 & $\cdot$ & $\cdot$ & $\cdot$ & $\cdot$ & $\cdot$ & 1 & $\cdot$ & $\cdot$ \\
\multicolumn{1}{|l|}{} & $\cdot$ & $\cdot$ & 1 & $\cdot$ & $\cdot$ & $\cdot$ & $\cdot$ & $\cdot$ & 1 & $\cdot$ \\
\multicolumn{1}{|l|}{} & $\cdot$ & $\cdot$ & $\cdot$ & 1 & $\cdot$ & $\cdot$ & $\cdot$ & $\cdot$ & $\cdot$ & 1 \\
\multicolumn{1}{|l|}{} & $\cdot$ & $\cdot$ & $\cdot$ & $\cdot$ & 1 & 1 & $\cdot$ & $\cdot$ & $\cdot$ & $\cdot$ \\ \hline

\multicolumn{1}{|l|}{\multirow{5}{*}{$x=3$}} & 1 & $\cdot$ & $\cdot$ & $\cdot$ & $\cdot$ & $\cdot$ & $\cdot$ & 1 & $\cdot$ & $\cdot$ \\
\multicolumn{1}{|l|}{} & $\cdot$ & 1 & $\cdot$ & $\cdot$ & $\cdot$ & $\cdot$ & $\cdot$ & $\cdot$ & 1 & $\cdot$ \\
\multicolumn{1}{|l|}{} & $\cdot$ & $\cdot$ & 1 & $\cdot$ & $\cdot$ & $\cdot$ & $\cdot$ & $\cdot$ & $\cdot$ & 1 \\
\multicolumn{1}{|l|}{} & $\cdot$ & $\cdot$ & $\cdot$ & 1 & $\cdot$ & 1 & $\cdot$ & $\cdot$ & $\cdot$ & $\cdot$ \\
\multicolumn{1}{|l|}{} & $\cdot$ & $\cdot$ & $\cdot$ & $\cdot$ & 1 & $\cdot$ & 1 & $\cdot$ & $\cdot$ & $\cdot$ \\ \hline

\multicolumn{1}{|l|}{\multirow{5}{*}{$x=4$}} & 1 & $\cdot$ & $\cdot$ & $\cdot$ & $\cdot$ & $\cdot$ & $\cdot$ & $\cdot$ & 1 & $\cdot$ \\
\multicolumn{1}{|l|}{} & $\cdot$ & 1 & $\cdot$ & $\cdot$ & $\cdot$ & $\cdot$ & $\cdot$ & $\cdot$ & $\cdot$ & 1 \\
\multicolumn{1}{|l|}{} & $\cdot$ & $\cdot$ & 1 & $\cdot$ & $\cdot$ & 1 & $\cdot$ & $\cdot$ & $\cdot$ & $\cdot$ \\
\multicolumn{1}{|l|}{} & $\cdot$ & $\cdot$ & $\cdot$ & 1 & $\cdot$ & $\cdot$ & 1 & $\cdot$ & $\cdot$ & $\cdot$ \\
\multicolumn{1}{|l|}{} & $\cdot$ & $\cdot$ & $\cdot$ & $\cdot$ & 1 & $\cdot$ & $\cdot$ & 1 & $\cdot$ & $\cdot$ \\ \hline

\multicolumn{1}{|l|}{\multirow{5}{*}{$x=5$}} & 1 & $\cdot$ & $\cdot$ & $\cdot$ & $\cdot$ & $\cdot$ & $\cdot$ & $\cdot$ & $\cdot$ & 1 \\
\multicolumn{1}{|l|}{} & $\cdot$ & 1 & $\cdot$ & $\cdot$ & $\cdot$ & 1 & $\cdot$ & $\cdot$ & $\cdot$ & $\cdot$ \\
\multicolumn{1}{|l|}{} & $\cdot$ & $\cdot$ & 1 & $\cdot$ & $\cdot$ & $\cdot$ & 1 & $\cdot$ & $\cdot$ & $\cdot$ \\
\multicolumn{1}{|l|}{} & $\cdot$ & $\cdot$ & $\cdot$ & 1 & $\cdot$ & $\cdot$ & $\cdot$ & 1 & $\cdot$ & $\cdot$ \\
\multicolumn{1}{|l|}{} & $\cdot$ & $\cdot$ & $\cdot$ & $\cdot$ & 1 & $\cdot$ & $\cdot$ & $\cdot$ & 1 & $\cdot$ \\ \hline

\end{tabular}
\caption{The coefficients of the Bell inequality in $(5,2,5,5)$. The table is written in full probability notation with the dots denoting zeros. The polytope bounds are $S_{\mL}=6$ and $S_{\mC}=7$. The maximally entangled bipartite five-dimensional states violate this with at least a score of $S_{\mQ}=7.1777$.}
\label{tab:5255}
\end{table}

\revise{Our Bell inequality is} a generalization of a previous correlation found in Ref.~\cite{sidajaya2023neural} to the $(5,2,5,5)$ scenario, shown in Table \ref{tab:5255}. This game is a truncated version of the XOR-$d$ games for $d=5$ \cite{buhrman2005causality,ji2008multisetting,liang2009reexamination}. Up to some relabeling, the XOR-$d$ games can be written as
\begin{align}
    \mathcal{I}_d(\mathcal{P}) = \sum_{a,b,x,y}^{d-1} \left[a - b = xy \; \text{mod} \; d\right] P(a,b|x,y),
\end{align}
where $[\cdot]$ is the Iverson bracket, which evaluates to 1 when the expression inside is true and 0 otherwise. While $a,b,x,y$ usually runs from $\{0,1,\dots,d-1\}$, here we truncate $b$ to $\{0,1\}$.

This Bell inequality has local bound $S_{\mL}=6$ and no-signaling bound $S_{\mNS}=10$. More importantly, the 1 bit bound is $S_{\mC}=7$ while the quantum bound is in the range of $7.1777 \leq S_{\mQ} \leq 7.1788$, where the lower bound is obtained by an explicit example and the upper bound by the level 1+AB NPA hierarchy \cite{navascues2008convergent}. The NPA hierarchy gives a series of relaxations of $\mQ$ that approaches $\mQ$ as the level gets higher and higher. Our upper bound is obtained by the level 1+AB NPA hierarchy that is implemented with Ref. \cite{johnston2016qetlab}. The total dimension of the probability space of this scenario is 250, signifying an improvement from the previously reported smallest example for fixed directional communication: The (truncated, asymmetrized) two magic squares found in Ref.~\cite{marton2023beating} by the method of parallel repetition feature a total dimension of 5376.

This can be further generalized to higher $d$. For example, the $d=6$ case in $(6,2,6,6)$ yields $S_{\mL}=7$ , $S_{\mC}=8$, and $S_{\mNS}=12$. The maximally entangled bipartite six-dimensional states violate this with a score of $8.3173<S_{\mQ}<8.3693$ (level 1+AB NPA). In general, for a $(d,2,d,d)$ scenario, $S_{\mL}=d+1$, $S_{\mC}=d+2$, and $S_{\mNS} = 2d$. 

\subsection{Violating the inequality with quantum mechanics}

\begin{figure}
    \centering
    \includegraphics[width=\linewidth]{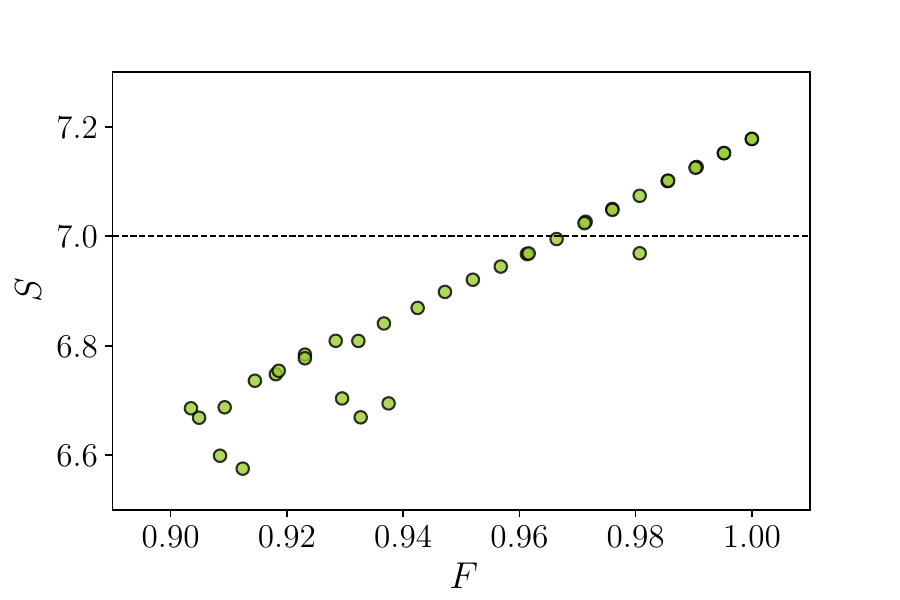}
    \caption{The plot shows the robustness of the violation against noise in the state. For each data point, we add Gaussian noise to the entries of $\rho = \ket{\Psi}\bra{\Psi}$ to get a noisy state and optimize the measurements numerically to obtain the maximum quantum score for that state. Since the optimization is done heuristically, the points represent only a lower bound and sometimes the optimization fails to converge on the global maximum, which gives the outliers. The fidelity of the noisy state ($F$) is the $x$-axis, while the score ($S$) is the $y$-axis. The dotted line is the local bound $S_\mL$.}
    \label{fig:1}
\end{figure}

The lower bounds \revise{of $S_\mQ$} were obtained by numerical optimization using MATLAB\revise{, which gave us the violating quantum correlations}. All were obtained by fixing the state to the maximally entangled state
\begin{align}
    \ket{\Psi} = \frac{1}{\sqrt{d}}\sum_{i=0}^{d-1}\ket{ii},
\end{align}
where $d$ is the dimension of the output. The \revise{corresponding} measurements are obtained numerically (link attached at the end), but we can still infer some patterns. Note that due to the nonlinear and heuristic nature of the optimization, the precision and optimality of the resulting measurements are significantly limited. In the following inferences we will use the symbol $\simeq$ to denote a rough equality. In most cases, this rough equality is satisfied within a 10\% margin, but in some cases, especially when the values are close to zero, this margin is not satisfied.

Let $\ket{a_j^i}\bra{a_j^i}$ $\Pi_j^i$ be Alice's projector that corresponds to the $i$-th measurement and $j$-th output. The first observation is that, $\forall i \in \mX $ and $\forall j,k \in \mA$,
\begin{align}
    \left|\braket{a_j^{i}}{a_k^{(i+1)\text{mod}d}}\right|^2 \simeq \left|\braket{a_j^{(i+1)\text{mod}d}}{a_k^{(i+2)\text{mod}d}}\right|^2.
\end{align}
That is, the inner products between the $j$-th and $k$-th basis vectors from two ``neighboring'' answers are always the same for every pair of neighboring answers. Second, Bob's two measurements form a mutually unbiased bases (MUB) with each other:
\begin{align}
    \left|\braket{b_i^{1}}{b^{2}_j}\right|^2\simeq \frac{1}{d}. \quad \forall i,j \in \mB.
\end{align}
Finally, the resulting behavior is a convex combination of the nonlocal behavior that maximally violates the inequality and white noise.
\begin{align}
    P(a,b|x,y) \simeq w P_\mNS(a,b|x,y)+ (1-w) \frac{1}{d^2},
\end{align}
where $P_{\mNS}(a,b|x,y) = V(a,b|x,y)/d$, i.e., Table~\ref{tab:5255} is reinterpreted as a nonsignaling behavior. For $d=5$, $w\simeq0.64$, while for $d=6$, $w\simeq0.63$.

Figure~\ref{fig:1} shows the heuristic quantum scores obtained by numerical optimization for noisy states in $d=5$. A fidelity of around $\sim 0.97$ is required to observe a violation. This presents a possibility for a challenging, but doable, implementation using a high-dimensional two-qudit state, such as those created in photonics, either using orbital angular momentum (OAM) \cite{mair2001entanglement,leach2010quantum,dada2011experimental,krenn2014generation,erhard2018twisted}, path \cite{rossi2009multipath, schaeff2012scalable, schaeff2015experimental, wang2018multidimensional}, time-bin \cite{stucki2005fabry, ikuta2016enhanced}, frequency-bin \cite{bernhard2013shaping,jin2016simple}, or time-energy encoding \cite{thew2004bell, richart2012experimental}. Recently, higher-dimensional entanglement has even been demonstrated outside of photonics using trapped ions \cite{hrmo2023native}.

\section{Conclusions}
Here we have presented a family of quantum correlations for $d\geq5$ which can not be simulated with a fixed directional communication from Alice to Bob with just a single bit of information. The $d=5$ case is, so far, the smallest example, both in the dimension of the quantum system and the probability dimension, of a quantum correlation that cannot be simulated with a single bit of fixed directional communication from Bob. Indeed, even if Ref.~\cite{marton2023beating} has examples with smaller output alphabets, the quantum realizations require higher-dimensional states. The small Hilbert space of this correlation presents a practical opportunity for an experimental implementation.

Our example can be extended for bidirectional communication by doing two of the games in parallel with reversed rules of Alice and Bob in the second game, resulting in a $(10,10,25,25)$ game. However, this does not improve on the two magic squares $(7,7,16,16)$ presented in Ref.~\cite{marton2023beating}.

The measurements to obtain the lower bounds of $S_\mQ$ can be found online \footnote{\href{https://github.com/PeterSidajaya/1 bit-part-deux}{https://github.com/PeterSidajaya/1 bit-part-deux}}.

\section*{Acknowledgements}
This research is supported by the National Research Foundation, Singapore and A*STAR under its CQT Bridging Grant. P.S.~thanks Yeong-Cherng Liang, Swati Pandey, Kai-Siang Chen, and Gelo Noel M.~Tabia for discussions and running some NPA hierarchy.

\bibliography{ref}% Produces the bibliography via BibTeX.

\end{document}